# TERRESTRIAL CONSEQUENCES OF SPECTRAL AND TEMPORAL VARIABILITY IN IONIZING PHOTON EVENTS

Larissa M. Ejzak[1,3], Adrian L. Melott[1], Mikhail V. Medvedev[1], and Brian C. Thomas[2]

ABSTRACT

Gamma-Ray Bursts (GRBs) directed at Earth from within a few *kpc* may have damaged the biosphere, primarily though changes in atmospheric chemistry which admit greatly increased Solar UV. However, GRBs are highly variable in spectrum and duration. Recent observations indicate that short (~0.1 s) burst GRBs, which have harder spectra, may be sufficiently abundant at low redshift that they may offer an additional significant effect. A much longer timescale is associated with shock breakout luminosity observed in the soft X-ray (~$10^3$ s) and UV (~$10^5$ s) emission, and radioactive decay gamma-ray line radiation emitted during the light curve phase of supernovae (~$10^7$ s). Here we generalize our atmospheric computations to include a broad range of peak photon energies and investigate the effect of burst duration while holding total fluence and other parameters constant. The results can be used to estimate the probable impact of various kinds of ionizing events (such as short GRBs, X-ray flashes, supernovae) upon the Earth's atmosphere. We find that the ultimate intensity of atmospheric effects varies only slightly with burst duration from $10^{-1}$ *s* to $10^8$ *s*. Therefore, the effect of many astrophysical events causing atmospheric ionization can be approximated without including time development. Detailed modeling requires specification of the season and latitude of the event. Harder photon spectra produce greater atmospheric effects for spectra with peaks up to about 20 *MeV*, because of greater penetration into the stratosphere.

[1] University of Kansas, Department of Physics and Astronomy, 1251 Wescoe Dr. # 1082, Lawrence, KS 66045-7582; ejzak@wisc.edu, melott@ku.edu, medvedev@ku.edu
[2] Washburn University, Department of Physics and Astronomy, Topeka, KS 66621; brian.thomas@washburn.edu
[3] presently at Department of Physics, University of Wisconsin

1. INTRODUCTION

Thorsett (1995) first pointed out the possible devastating impact of a gamma-ray burst (GRB) from within our Galaxy upon the Earth. Scalo and Wheeler (2002) refined this expectation and made quantitative estimates of effects. Smith, Scalo, and Wheeler (2004) did an explicit radiative transfer calculation of effects upon the biosphere, as well as upon Mars-like planets. More recently, we have proposed a possible connection with a particular mass extinction event (Melott et al. 2004) and computed in some detail the long-term atmospheric effects and some biogeophotochemical consequences of a "typical" long-duration GRB which have softer spectra (LSB) beamed at the Earth from 2 kpc, a conservatively estimated probable nearest event in the last Gy (Thomas et al. 2005a,b; Melott et al. 2005).

The greatest effect of the GRB photons is to make changes in atmospheric chemistry which generate odd nitrogen compounds ($NO_y$ = N, NO, $NO_2$, $NO_3$, $N_2O_5$, $HNO_3$, $HO_2NO_2$, $ClONO_2$, $BrONO_2$). Those usually called $NO_x$ (NO, $NO_2$) catalyze the depletion of ozone ($O_3$). $O_3$ normally prevents about 90% of the Solar UVB (280-315 nm) radiation from reaching the surface. UVB is extremely damaging to most organisms, particularly since it easily damages DNA as well as proteins. In Thomas et al. (2005b) effects were explored as a function of altitude, latitude and time for a burst at a variety of fluences, burst seasons and latitudes. Season and latitude of the burst event were found to considerably modulate the intensity of the atmospheric effects.

In that work the Band et al. (1993) GRB spectrum was used with a peak at 187.5 *keV* and duration of 10 *s*, typical LSB values. However, GRBs show both spectral and temporal variation. Burst duration ranges over several orders of magnitude, with an apparently entirely different class of short bursts, often of order 0.1 *s*. The peak of the Band spectrum shows considerable variation, and short burst GRBs often have a "harder" spectrum, with considerable flux above several *MeV* (Cline et al 2005); henceforth we shall refer to this class as short-hard bursts, or SHB. The SHB class may well comprise two or more populations, one cosmological and one in "nearby" galaxies. Although long-duration LSB rates have clearly been declining with cosmic time, there many other kinds of objects, such as "short-hard bursts", supernovae, soft X-ray flashes, etc. which might have terrestrial impacts. There is good evidence for at least one very young GRB remnant in our Galaxy (Atoyan et al. 2006). For further discussion see Dermer & Atoyan (2006).

It is important as well as to consider the spectral variation, luminosity evolution, and temporal variability of LSBs, including possible effects of afterglow on an already-damaged terrestrial atmosphere. It is impossible to fully explore the large parameter space relevant to effects on the Earth. In this paper, we explore the consequences of burst duration and photon energy, while holding other variables constant, in order to arrive at a more general understanding. As better rate data become available, this will help to clarify the relative effects of SHBs as well as soft gamma repeaters (e.g. Gotz et al. 2005), post-burst flares from GRBs (e.g. Falcone et al. 2006), X-ray flashes from supernovae (e.g. Campana et al. 2006; see also Scalo et al. 2001), possible large solar flares (Schaefer et al. 2000), long-duration sources like Cygnus X-3, radioactive decay gamma-ray line radiation emitted for ~ $10^7$ s during the light curve phase of supernovae (e.g. Teegarden et al. 1989, Karam 2002), and other possible sources of high-energy photons upon the Earth and Earthlike planets. As our results are computed based on the fluence at the Earth, they can be interpreted in terms of a variety of source energies and distances, and used to estimate the likely effects of various classes of events. Of course, many of these are very low-probability events, but we include a study of a wide variety of photon energy and duration to understand the dependence of the atmospheric response on these parameters.

Gehrels et al. (2003) studied the effect of relatively nearby supernovae on the Earth, using a different spectral model. While for extremely nearby (~8 *pc*) supernovae the effects are strong, it is difficult to disentangle the trends due to various assumed spectra, durations, etc. For this reason we include in this study events of very long duration (up to $10^8$ s), while keeping the fluence and spectrum constant. In this way we can isolate the effects of burst duration and spectral variation from others in their impact upon the atmosphere, which will help to understand the effects of nearby supernovae. In future work we will model more realistic supernovae.

2. METHODS

2.1 *Atmospheric model*
The atmospheric code used here is a 2D (latitude, pressure) time-dependent model with 18 equal latitude bins and 58 log-pressure bins, corresponding to approximately 2 *km* spacing. The code uses a lookup table for photolysis; mixing and small-scale winds are empirical. We will not repeat here a much more extensive description contained in Thomas et al. (2005b), including the Appendices and Tables. The code will be valid for atmospheric composition similar to the present Earth, which should be true for about the last 500 My, except for an interlude of very high oxygen which ended at the Permian extinction (Berner et al. 2003). There will also be differences in transport due to changes in continent placement, etc. As the present Earth is highly north-south asymmetric, the importance of this effect has been tested (Thomas et al 2005b) by the north-south symmetry of the results obtained; it is known that systematic differences due to continent placement, wind currents, etc. are small, particularly compared with effectively random effects of latitude and time of year of a given burst.

It is important to note that the chemistry is handled in bulk; once the overall ionization is computed, 1.25 $NO_x$ per ion pair produced; $NO_x$ is a "family" of compounds whose ratios are empirically based. No ions are included as intermediaries; their effects are subsumed in the bulk production rate. Some additional effects might come from the effects of intermediaries such as $N_2^+$. It has been argued (Jackman et al. 1979) that this ion might boost $NO_y$ production substantially above about 80 km, which is essentially irrelevant to the question of $O_3$ depletion, and that below this the effects are probably negligible.

The code has been used to compute effects on the atmosphere of ionizing radiation due to recent solar flares (Jackman et al. 1996 and references therein) and possible nearby supernovae (Gehrels et al. 2003). As described in Thomas et al (2005b), we use a version with a shortened timestep in order to account for the rapid variability of GRBs; here we have had to shorten the timestep even more to deal with SHB. The shortened timestep model was tested for consistency with earlier computations of the same burst duration. As before, after the burst, the values are input into a longer-term model with a one-day timestep which includes diffusion and transport.

The code does not self-consistently calculate heating effects, but we have estimated these not to exceed 10 °K both for direct heating by the burst and subsequent changes in stratospheric temperature due to lack of UV absorption by $O_3$, so these effects should be small. More discussion of heating effects due to $O_3$ changes is presented in Thomas et al (2005b).

2.2 *Burst models and input parameters*
We have controls in order to disentangle and clarify the effects of varying spectral and temporal properties of the burst. We use a burst at 45° N latitude at noon on the March equinox, which is near the median of resulting intensity of effects in the (season, latitude) parameter space. The total fluence is 100 *kJ m$^{-2}$*, which corresponded to the expected nearest LSB beamed at the Earth within the past *Gy* based on rate estimates (Melott et al. 2004; Thomas et al. 2005b), though Dermer & Holmes (2005) estimate a somewhat closer event. There is of course considerable uncertainty in the fluence due to uncertainties in the low-redshift rate in galaxies like the Milky Way, as well as due to the statistics of random events. Note that varying the fluence in Thomas et al (2005b) showed effects changing less than linearly; ozone depletion there could be crudely characterized as varying with the cube root of the fluence. Therefore these uncertainties in fluence translate into less uncertainty in the likely effect upon terrestrial planets. For example, using the thin disk galaxy approximation, events of fluence 10 *kJ m$^{-2}$* would have ten times the effective surface area of origination, and be expected on average ten times as often as events with ten times more fluence. But their effect upon the biosphere might be crudely characterized as about half as severe (see Thomas et al. 2005b, particularly Figs. 17 & 18 for more detail). Thus, a much lower LSB event rate as advocated by some (Fruchter et al. 2006; Stanek et al. 2006) does not translate into proportionately lower rates of trauma to the biosphere.

We vary spectra and duration around this previously modeled, extensively studied choice of parameters.  Our previous work used the Band et al. (1993) spectrum with a peak at 187.5 *keV*, a typical long-duration GRB value.  One of our primary goals here is to vary the peak energy.  We do this by continuing to use the same spectral fit, but varying the peak energy by one order of magnitude over steps 1.875, 18.75, 187.5 *keV* and 1.875, 18.75, 187.5 *MeV*.  As before the parameters in the Band fit are α = -0.8, β = -2.3 (Preece et al. 2000).  This spectral shape was used previously as a good fit to a typical LSB; it is now used solely to avoid varying more than one parameter as we investigate the effect of photon energy.  While this spectral shape (See Fig. 1—note we plot energy flux not photons) is somewhat arbitrary when considering events other than LSB, any other shape would be equally arbitrary for generic events.

As we are considering considerably harder and softer spectra than before, a wider range of attenuation coefficients are needed than those present in the lookup table embedded in the code.  We have used the NIST XCOM database, available online (Berger et al. 2005) based on a mixture of 79% N and 21% O, except for 100 eV to 1 keV, which is not present there.  In this range we use the data from Plechaty et al. (1981). Using this, we are able to include effects of photons impacting the upper atmosphere with initial energies from 100 eV to 100 GeV.  The atmospheric code includes more constituents, but their abundance is small enough not to affect the attenuation of high-energy photons.

Some limitations should be noted. The cross sections for elements in the XCOM database pertain to isolated neutral atoms, and do not take into account molecular effects which modify the cross sections, especially in the vicinity of absorption edges. Relatively small cross sections, such as those for Delbruck scattering, two-photon Compton scattering or photo-meson production, are not included. Also omitted is the nuclear absorption and re-emission which, in the giant-dipole resonance region from 5 MeV to 30 MeV, can contribute a few percent to the total attenuation coefficient.  However, pair production in the nuclear field, which can dominate above 30 MeV, is fully included (see, e.g. Groom 1998)  Finally, XCOM does not calculate energy absorption coefficients that represent the conversion of photon energy to kinetic energy of secondary Compton-, photo-, and pair-electrons.

Though XCOM provides opacities/absorption coefficients for photons of a wide range of energies, propagation and Compton scattering at high energies, above 10 MeV or so, must be treated with care and should involve detailed radiative transfer calculations. The problem arises from the Klein-Nishina correction to Compton scattering at large energies. The scattering cross-section is suppressed roughly as ~1/E at large scattering angles. Thus, high-energy photons will be predominantly scattered in the forward or backward direction, which makes use of angle averaged cross sections somewhat inaccurate in stratified atmospheres. Accurate radiative transfer computations have been done at lower energies by Smith, Scalo & Wheeler, (2004). Implementation of similar computational techniques in the existing atmospheric code is not feasible at present. However, since the high energy mass attenuation coefficients we use are checked by laboratory experiment, and this approximate treatment should not be off by more than few percent (Hubbell et al. 1980), the results should be sufficiently accurate, and average over these effects. Since the cross-section is fairly flat above a few MeV, we do not expect large changes with increasing energy beyond this level. However, this expectation needs to be checked.

GRBs exhibit considerable diversity of duration for the prompt emission, though in our earlier work we used a fiducial 10 *s* burst. Also, many other kinds of ionizing events are possible, with new candidates appearing at least yearly. For these reasons it is important to understand the effect of energy incident on other timescales. The clearest and most straightforward way to accomplish this is to consider the same fluence delivered over other times. We have therefore examined the effect of our original 187.5 *keV* peak spectrum, with the total fluence delivered over 0.1, 10, $10^3$, $10^5$, $10^6$, $10^7$, and $10^8$ *s* which more than covers the observed range of extrasolar high-energy sources observed from SHB through supernovae. We had previously shortened the timestep of the initial burst code from 225 *s* used for solar flares (Jackman et al. 2001) to 1 *s*; we now shorten it to 0.01s for the 1s and 0.1s burst models, with consistency checks implemented. It is important to note that while varying the spectrum and duration of the bursts individually, we have held the total fluence at the Earth constant at 100 *kJ* $m^{-2}$. Our goal is not to model every possible type of event, but to unearth general principles that will help to understand effects and estimate the effect of various phenomenae, some of which have yet to be observed. We have not modeled events with greater fluence. They may happen, but exceed the likely range of validity of our atmospheric code, due to temperature and transport feedback effects (Thomas et al. 2005b). In the present case heating of the stratosphere due to the prompt flux is limited to about 10 °K, and later temperature changes due to changes in the heating of the stratosphere from changed ozone concentration are of similar magnitude. Therefore, these effects, which are not included in the code, are small. Fortunately, events with fluences above 100 *kJ* $m^{-2}$ which we cannot reliably model, are, given present knowledge, improbable at the Earth on Gy timescales.

GRBs may produce a substantial flux of very high energy cosmic rays (e.g. Waxman 2004, Dermer & Holmes 2005), and a nearby supernova certainly would. This could produce a substantial additional effect beyond that of the photons, modeled here. The size of the expected CR flux is uncertain, and its propagation in the galactic magnetic field and the Earth's magnetosphere is strongly energy-dependent, and could be impulsive or diffusive (depending upon CR energy and characteristics of the local galactic magnetic field) to the target planet. For these reasons we do not include this, though it clearly could have an impact comparable to that of the photons, for increased stress on the biosphere.

3. RESULTS

3.1 *The unperturbed ozone layer*
There are at least three possible effects on biota, discussed in Melott et al. (2005): increased UV due to depletion of the protective ozone layer, possible climate change caused by reduced sunlight due to absorption by $NO_2$ produced in the atmosphere, and nitrate deposition as it is removed by rainout in the form of dilute nitric acid (in order of estimated declining importance). These are all closely coupled; we will describe results here ultimately in terms of the DNA damage due to solar UVB transmitted through the $O_3$-depleted atmosphere. This appears to be the dominant effect for terrestrial planets, though direct irradiation from the event itself may dominate for Mars-like and other planets with thin atmospheres (Smith et al. 2004). Understanding this effect requires attention to the structure of the normal atmosphere. In Fig. 2, we show the structure of the ozone layer output from our unperturbed atmospheric model. This reproduces the structure of the atmosphere well, except for effects of anthropogenic compounds such as chlorofluorocarbons which we have removed from the code. Normally, $O_3$ is produced as a side effect of photolysis near the equator and at high altitude; it is transported to the poles where it tends to accumulate. Some asymmetry of latitude is visible in the plot, due to the present configuration of continents on the Earth as well as the latitude and time of year. The first of these is a limitation of our model when used for paleontological purposes, but is known to be small compared with the latter two causes mentioned (Thomas et al. 2005b).

3.2 *Effect of incident ionizing radiation*
For photons over the wide range of energies we shall consider, the attenuation coefficients (Berger et al. 2005) are such that only a tiny fraction of the incident radiation would reach the ground for a terrestrial-type atmosphere (Smith et. al 2004). Nearly all the energy goes into atmospheric chemistry. The primary chemical effect of the incident radiation is to break the strong chemical bonds of $O_2$ and $N_2$, making possible the formation of molecules which are normally present in very low abundances in the atmosphere. NO and $NO_2$ are in this class; they also catalyze the destruction of ozone. We have found that it takes nearly a decade for the atmosphere to recover from a burst as they are rained out and things return to the former nearly steady-state. $O_3$ normally shields the Earth from most of the solar UVB (280-315 *nm*). During the time $O_3$ is depleted, organisms on the surface will be exposed to drastically increased levels of UVB, which is normally about 90% absorbed by ozone. UVB is strongly absorbed by DNA, breaking chemical bonds and leading to cancer and mutations. UVB levels found as a consequence of previous burst models are far in excess of those known to cause mortality in marine organisms, as summarized in Thomas et al (2005b). UVA (315-400 nm) is also known to cause significant biological damage but can also be helpful in photosynthesis, which complicates things. However, UVA levels will not be changed noticeably by $O_3$ depletion, but may be somewhat lowered by $NO_2$ absorption.  The relative contribution of UVA and UVB to biological damage is controversial and varies with the kind of effect considered.

In order to induce this effect, the $NO_y$ compounds must reach the ozone; one of the questions answered by our previous work was whether the photons would reach deeply enough into the atmosphere to do so.  They do for a typical LSB; but since cross-sections vary with energy, this is something that may strongly modify the effect of other kinds of energetic events.

3.3 *Odd nitrogen and ozone production*
Energy from a burst could reach the ground and be a modest threat for intense nearby events, but this is dwarfed by the long-term effects of solar UVB for terrestrial type atmospheres.  Nearly all of the energy goes into ionizing, dissociating, and heating the atmosphere from the stratosphere on up.  The greatest departure from the usual state is the breaking of the extremely strong $N_2$ triple bond.  This bond strength is the reason organisms need to "fix" nitrogen to make it available, and it is the reason that the atmosphere does not normally contain a large abundance of odd nitrogen compounds.   However, the irradiation we are considering will break this bond as well as that of $O_2$. The result is shown in Fig. 3, which is one day after switchover to the one-day timestep code, or five days after the burst (however, $NO_y$ densities are almost the same one week later).  Plots of $NO_y$ as a function of latitude and log pressure (appx altitude) are shown for the six peak photon energies from 1.875 *keV* to 187.5 *Mev*.  Again, the total fluence is the same in all cases.  The difference in results is striking—effects are much greater for harder spectra.  This can be well understood as an effect of the attenuation lengths.  Softer bursts interact much higher in the atmosphere. Note by comparing Figs 2 and 3 that there is substantial overlap between the vertical distributions of ozone and $NO_y$, even for our lowest energy photons, which increases with photon energy. Thus, the increase in effect with increased photon energy is not many orders of magnitude, although it is substantial. Reaction rates are different at different pressure, but most importantly the destruction of $NO_y$ compounds by photolysis goes on more rapidly at higher altitude, where solar UV is much more intense. (In our one week plots, not shown, the only visible difference from one day is some reduction of $NO_y$ by photolysis at high altitude).  At lower altitude, the unperturbed atmosphere is largely self-shielded against UVC and UVB.   Ironically, existing ozone helps to shield the odd nitrogen which will eventually catalyze its destruction. When upper atmosphere ozone is depleted, there is some increased synthesis of ozone in the lower atmosphere.

These results also serve as a check on the Thomas et al (2005b) results, which contained attenuation data up to 10 *MeV*, and assumed a spectral cutoff there, but the same total fluence.  With the extended spectrum here, but the same (187.5 *keV*) peak energy, effects are only a few percent greater.

To summarize, the primary effect of increasingly hard incident spectra is that the photons penetrate much more deeply into the stratosphere, where they are able to produce a great deal more odd nitrogen compounds, coincident with the location of the solar UVB-shielding ozone. As we will discuss later, transport may redistribute the $NO_y$ compounds, but is far too slow to compete in a major way with any processes of their creation or effect on ozone for many months.

3.4 *Ozone depletion*
In previous work (Melott et al 2005) we described three effects of potential importance to a terrestrial-type biosphere from an ionizing radiation burst. They are increased UVB due to ozone depletion, increased opacity in the visible due to $NO_2$ production which might reduce surface temperatures, and nitric acid deposition at the surface. For burst fluences probable over geologic timescales, the ozone depletion is clearly of major importance, while the others are questionable. For this reason, we describe the ozone depletion effect; the others scale closely with it.

Fig 4a shows fractional change in ozone density one week after the onset of the one-day timestep code, or 11 days post-burst. Much of the variation with incident spectral hardness is understandable as a consequence of the odd nitrogen production. For harder spectra, ozone depletion extends deeper into the stratosphere. There is also some ozone production in the upper troposphere and lower stratosphere, shown as a positive percentage change. This production is not a large mass, as ozone is initially not of high abundance at low altitudes. It results from the small amount of incident burst ionizing radiation that penetrates that far, combined with an enhanced solar flux. Note that our print version has a bimodal gray scale; both the ozone depletion at high altitude and the ozone production at lower altitude and latitude are shown as darkened areas for this Figure.

Since percent change is plotted here, the biggest absolute effect is the depletion of the stratospheric ozone. The mass of ozone lost from the upper stratosphere exceeds the mass of ozone produced in lower stratosphere. The log-pressure, or approximately altitude axis is somewhat visually misleading as to the size of the effect. For this reason in Fig 4b we use a linear pressure (appx log -altitude) axis. It is clear here that the vertical mass of ozone remaining decreases strongly with spectral hardness as the peak moves up to about 20 *MeV*

The time evolution of the global average ozone column density is shown for all our burst spectra in Fig 5. The lines are not labeled, as the trend is simply increased depletion for increasing peak photon energy. Again, as a check on the Thomas et al. (2005b) results, we find close to 33% peak mean column density for the (187.5 *keV*) case in which about 30% was found there. The difference can be understood as our inclusion of the photons in the power-law beyond 10 *MeV*, which penetrate the stratosphere.

The increase in ozone depletion with energy as plotted here includes a partially compensating effect: the increased synthesis of ozone at lower altitudes made possible by greater depletion (shown in Figs 4). We believe it is relevant to include this lower-altitude ozone, as it does contribute to the shielding column density against UVB at the surface.

Changes over the first few months in the lowest-energy photon cases are a consequence of the higher altitude of energy deposition. In these cases, photolysis of the $NO_y$ products at high altitude results in their decline and a partial ozone recovery at these altitudes. On the other hand, in the highest energy cases ozone depletion continues to worsen at for several weeks. In these cases there is less energy deposition at high altitudes, but some reaches the lower stratosphere. NO densities of order $10^8$ at these altitudes lead to an ozone destruction timescale of days. While the NO is not noticeably transported, ozone depletion can be seen (not shown) to worsen in the lower stratosphere, most strongly for the higher energy cases. The rate of this slow ozone depletion is (at high latitudes and low altitudes) competitive with the new rate of ozone synthesis, whereas at low latitudes, with a higher Sun angle, synthesis dominates at lower altitudes. The photolysis loss of $NO_y$ and slight ozone recovery dominates overall below the 187.5 keV energy; completion of the reaction and slight worsening of ozone depletion dominates overall above 187.5 keV. We do not find sufficient ozone synthesis for toxic effects.

Values for the vertical eddy diffusion coefficient ($K_z$) in the model are from Fleming et al. (1999). There is uncertainty in this quantity which may affect our results. We are primarily concerned with how this transport may move $NO_y$ produced at higher altitudes (ie. in lower peak energy bursts) down to where the greatest concentrations of $O_3$ occur.

The value of $K_z$ varies from 0.2 $m^2\,s^{-1}$ at the bottom of the stratosphere to about 0.02 $m^2\,s^{-1}$ between approximately 20 and 30 km altitude and then climbs again toward the lower stratosphere value (0.2 $m^2\,s^{-1}$) going upward (to about 40 km). The exact value is dependent on seasonal variations as well as latitude. For this discussion we will use the annual average at the equator.

The distance a constituent is moved by this diffusion is given by $(K_z\,t)^{1/2}$. Taking $K_z = 0.15\,m^2\,s^{-1}$ (the value at about 40 km) and t = 1 day, the distance is 113 m. If the actual $K_z$ value should be a factor of 10 larger, then this distance is about 360 m. Therefore, this process is not a major factor in moving $NO_y$ over tens of km on timescales shorter than weeks to months. In addition, even a factor of 10 uncertainty does not greatly affect the results, particularly on short timescales. The more critical quantity to our results, not affected by this uncertainty, is transport across the boundary into the troposphere, where rainout can begin.

For all cases, the ozone depletion then remains approximately constant for about two years. Some recovery is noticeable after five years, and is effectively complete in a decade.  This is many generations for the highly UV-transparent single-celled organisms which lie at the base of the marine food chain.  Environmental consequences will be severe, and more so for events with increasingly hard spectra.

3.5 *DNA damage*
Depletion of atmospheric ozone will cause substantially more UVB to penetrate the atmosphere and irradiate organisms near the surface.  UVB interacts strongly with protein and DNA molecules, damaging them.  This can cause gross damage, as well as cancers and an increase in the mutation rate.  Deep water organisms should be affected only indirectly, as UVB photons are absorbed by a few meters of water; however, they may suffer to the extent that most are dependent on a food chain that begins on and near the surface, with phytoplankton.  There are a variety of ways to measure biological damage, as discussed in Thomas et al. (2005b).  "Action spectra" are defined for various biological effects.  The strong effects of ozone depletion are for those that peak in the UVB regime.  There is increasing appreciation of the damaging effects of UVA, which may be competitive with UVB damage, and even some evidence for DNA damage in Hamster cells with blue light (Kielbassa et al. 1997).  On the other hand, UVA is beneficial for some photosynthetic organisms (Mangelt & Prezelin 2005).  Only UVB will be substantially enhanced as a result of the $O_3$ depletion.   In fact, UVA and blue/violet may be slightly reduced due to the opacity of brown $NO_2$ in this band.  Since UVB will increase greatly, and UVA/violet/blue will be constant or slightly decreased, it is appropriate to focus on the UVB damage. We follow Thomas et al (2005b) in using the DNA damage action spectrum of Setlow (1974). This action spectrum is an average over many kinds of damage to E. Coli, DNA, phages, etc. which do not lie more than a factor 2 (usually much less) off Setlow's fitted curve.   We plot relative, not absolute effects, which reduces the sensitivity of our presentation to the action spectrum.  Our proxy for DNA biological damage is the UVB radiation transmitted by the depleted atmosphere per day, convolved with this biological action spectrum, taking into account the Sun angle, length of the day, etc.  For this purpose, we do a full computation of the changed optical depth due to changed $O_3$ in the post-burst atmosphere in 4 nm wavelength bands over the range 280-315 nm. The solar radiation which is transmitted through this depleted atmosphere is convolved with the Setlow action spectrum and integrated over time giving an illustrative UVB-based biological effect as a function of time and latitude.  Of course, overall mutation and cancer rates may vary nonlinearly with the assumed DNA damage rates, because of repair mechanisms which are not taken into account here.  But these plots should at least give a good approximate indication of the approximate increased damage to photosynthetic single-celled organisms which lie at the base of the marine food chain.

In Fig 6 we show the result of this computation as a function of time for a typical pre-burst year followed by the first three post-burst years.  There is considerable evidence (summarized in Melott et al. 2004, Thomas et al. 2005b) that increases of a few times 10% in this function can lead to severe damage and mortality for marine microorganisms. As expected, the results are more severe with increasing spectral hardness, up to 20 *MeV* peak. We are not now modeling any particular observed event, but we may compare these results with those noted in Thomas et al (2005b). An order of magnitude increase in fluence approximately doubled the ozone depletion in that study. We find here that an order of magnitude increase in photon energy substantially increases the amount of ozone depletion. For example, a 2 keV peak burst with fluence 100 kJ m$^{-2}$ could be expected to do about as much damage as a 200 keV peak burst with fluence10 kJ m$^{-2}$. There is a substantial differential in effect for harder spectra, which may partially compensate for lower luminosity or greater distance. Using the one year post-burst globally averaged ozone depletion values in Figure 5 to estimate the amount of UVB that would reach the surface compared with unperturbed values for the 100 *kJ m$^{-2}$* irradiation, one arrives at "average" values from 1.5 to 2.7 times normal. As can be seen, localized pre-burst values of the UVB flux under the biological weighting window never go much above twice the global mean, but post-burst they reach local values from approximately 4 to 15 times this mean, depending upon the spectrum. Since values of a few times 10% UVB increase are typically lethal for phytoplankton and similar organisms that lie at the base of the food chain, none of these are remotely "safe" for the biosphere. Although detailed modeling of individual events is needed, one can estimate an approximate level of effect by scaling from these results.

3.6 *Trends with burst duration*
Varying around our "standard" 100 *kJ m$^{-2}$* burst peaking at 187.5 *keV*, we have examined the ozone depletion history of seven simulations in which the energy is deposited over timescales ranging from 0.1 s to 10$^8$ s. The latter is longer than any plausible nearby event type (but might be relevant if an object such as Cygnus X-3 were within a few pc of a terrestrial planet). It is included to better elaborate trends. The following simple result may be the most useful in this paper: there is no major duration effect. The maximum global average ozone depletions range between 33% and 37%; there are no large differences in the duration of the effect or the time-integrated ozone deficit. This result makes possible evaluation of the atmospheric impact of ionizing photon events without detailed modeling of the history (e.g. GRB variability, supernova light curve) by simply using the integrated flux. To first order, the ozone depletion is limited by the manufacture of ionized and dissociated N and O which is limited by the incident fluence.

Let us first discuss the physical basis of this result. It says that the fluence, not the flux, is the best estimator of the size of the effect in the duration range we have considered. It implies that to first order the effect of one photon is largely independent of the effect of another photon. Why is this true, and why do rate-limiting reactions have so little effect? We will mention a number of general conditions, then move to some specifics.

First, the dissociation of $N_2$ by photons and subsequent formation of reactive compounds must be more rapid than the removal of these compounds. This is easily met; the timescale of the former is seconds to minutes, and of the latter (by transport to the troposphere and rainout, mostly as nitric acid) is many years. Of course, irradiation for periods longer than this would obviously compete with rainout. Secondly, reaction of N to form $NO_x$ must be faster than competing reactions, including photolysis. We find that photolysis is only important in the ionosphere and far upper stratosphere, above the $O_3$ layer, and that its timescale is days to weeks. Other reactions (see below) also cannot compete with this primary channel.

The primary driver to modify atmospheric chemistry is atomic N. This can recombine through the 2-body reaction

$$N + N \rightarrow N_2 + \gamma \tag{1}$$

This is so severely constrained by simultaneous momentum and energy conservation that the available phase space is very small. We have not found any laboratory measurements of this rate—it is so slow that even at very low pressures it is superceded by interactions with container walls or with trace contaminants. It may become important in the upper ionosphere, at altitudes too high to affect our ozone results. A more relevant competitor is the three-body interaction

$$N + N + M \rightarrow N_2 + M \tag{2}$$

where M is an arbitrary third body and may carry off kinetic energy. For the Earth's atmosphere, $N_2$ is the most plausible identity for M. Reactions for $N^+$ have similar timescales. Making the most pessimistic assumptions (all interacting photons are instantaneously converted to N atoms), we find that this reaction is still about 25 orders of magnitude slower than the most important short-term, *in situ* reaction for the recombination of N,

$$N + NO \rightarrow N_2 + O, \tag{3}$$

which is included in our code, does have some effect, and whose effects compete slightly with the creation of $O_3$-depleting compounds as discussed below. Once newly formed N is removed by this or other (less significant) reactions, it is out of the depletion cycle.

Most of the products of ionization and dissociation begin their ozone-depleting interaction with the atmosphere by forming NO. The most important reaction is

$$N + O_2 \rightarrow NO + O \tag{4}$$

$NO_2$ is easily formed from NO. We will not repeat here the $O_3$ depletion catalysis cycle, except to mention that it involves primarily NO and $NO_2$, which are recycled and not depleted by it.

We can discuss the relative importance of (3) and (4) given abundances in the normal or postburst atmosphere, and rate constants given in Thomas et al (2005b). The timescale for reaction (4) to proceed in the mid-stratosphere is of order 5s. Under normal (preburst) conditions, the timescale for (3) is of order 30s, so it has only a modest effect. The opposite extreme within our range is when the maximum NO concentration is reached, the highest single concentration region in any place in any of our models would cause the timescale of reaction (3) to drop to 0.2s.  However, a more typical postburst global value for would cause reaction (3) to eventually reach a timescale comparable to reaction (4), of order 5s. This introduces a rate-limiting, negative feedback effect as N is removed.  However, this rate-limiting reaction can only be significant after a great deal of NO is formed, by reaction (4), so it is less effective in our shortest bursts. It would also lose some effectiveness in our very longest duration cases: latitudinal transport over a period of months will move NO around, while N will still be formed primarily at higher altitudes and at latitudes which maximize the irradiation by the source.  When the NO and N are not in proximity, (3) will have much more difficulty in moving forward.  To summarize, reaction (3) can only compete with (4) for times that are long enough to allow significant NO to build up while the irradiation is on, but not so long that transport can move NO out of the region where the radiation is primarily absorbed.  So, for the most part, the effective recombination in (3) plays a minor role.

Ozone depletion timescales in place are fairly rapid, reaching local "equilibrium" values within minutes (but days in the lower stratosphere where $NO_y$ concentrations are lower, see discussion in Section 3.4).. Over a period of weeks reactions even in the lower stratosphere take place which maximizes ozone depletion with the available catalysts.  Therefore the $O_3$ depletion rather closely tracks the $NO_y$ abundance. Some small amount of $NO_y$ is destroyed by photolysis in the upper stratosphere and mesosphere.  The much more important and longer-term process for $NO_y$ destruction is downward transport into the troposphere and rainout as $HNO_3$, which begins within months and takes nearly a decade to complete. To summarize, most of the $NO_y$ which is not destroyed in the first few minutes by reaction (3) is likely to survive for months and years.

The long-term history of the global average ozone depletion for various burst durations is shown in Figure 7a, and its behavior on the first day for the shorter bursts is shown in Figure 7b.  As discussed previously, the reactions in the lower stratosphere can take days to roughly equilibriate, given lower $NO_y$ concentrations there.  This effect is of course not visible in those irradiation events which take place over a timescale longer than this time; in such cases the delay is dominated by the delayed arrival of all the photons.  Eventually they all reach approximately the same depletion.

There are of course some limits on the use of the fluence approximation; it is not absolute:  There are weak trends with duration. Overall ozone depletion is maximized for very short and very long bursts, with a plateau at $10^3$ -$10^6$ s. The trend for very short bursts is associated with the NO destruction timescale. During and soon after a burst, there is considerable atomic nitrogen present. This destroys NO through the reaction (3), as discussed above.  Our shortest bursts 0.1 s begin to deplete ozone as soon as their reaction products are formed

(usually beginning with reaction (4)), since many of the photons are captured within the stratosphere and their products do not need to be transported in order to begin damage.  Reaction (4) makes NO, so if that has completed, then there is more probability of reaction (3) taking place.  Therefore, 10 s bursts have some increased competition with the reaction (3), and at $10^3$ s or longer there is even more competition.  Therefore, these bursts deplete slightly less ozone. A plateau is reached.  Then, above $10^7$ s transport to the troposphere and rainout just begins to remove $NO_x$ from the atmosphere, and ozone depletion/recovery timescales come into play (see Figures 3 and 7 in Thomas et al. 2005b).  $NO_x$ compounds created over very extended time periods show a delayed and slightly intensified depletion effect, as evident in Fig 7b.  The delay is a straightforward consequence of delayed arrival of photons, and this is the main effect.  An NO molecule in a depleted atmosphere has a lower probability of resulting in additional ozone depletion.  But when the atmosphere has begun to recover, the photon has a higher probability of resulting in ozone depletion than if it had arrived earlier.  So, events longer than $10^7$ s keep the NO abundance higher at later times, and each NO molecule will have slightly increased effectiveness as compared with it being formed all about the same time in a short burst.  On a timescale of longer than a year, vertical mixing can become important through mixing between the stratosphere and the troposphere, and transport via the Hadley cells and related phenomenae, which cause the $10^8$ s irradiation to show a substantially increased depletion.  On even longer scales, rainout will dominate and decrease depletion.

## 4. DISCUSSION

### 4.1 *Effect of burst duration and temporal variability*

We have found that over many orders of magnitude, holding the total fluence constant, that varying the burst duration over the range of 0.1 to $10^8$ s had a small effect on the overall atmospheric consequences at the Earth.  To first order the ultimate level of ozone depletion was nearly the same, with onset delayed till all the photons had arrived. This is undoubtedly a consequence of the fact that once a photon ionizes and dissociates atmospheric molecules, there are a variety of reaction pathways largely independent of the ionization of other molecules, since the ionization fraction is never close to unity for our range of fluences, and the reaction products usually do not reach high enough abundances to substantially feed back onto their rates.  This was anticipated, based on a rough comparison of effects between Thomas et al (2005b) and Gehrels et al. (2003), which dealt with GRBs and supernovae respectively.  This suggests a simplifying approximation that ozone depletion effects depend primarily on the integrated fluence of an ionizing event.  We have seen some increased depletion for timescales at the very short (< 10s) and very long (> $10^7$ s) end of our range, but these variations are a few percent, comparable to (for example) day-night differences in the timing of a burst event, and smaller than variations due to latitude or season of the burst.  There are strong short-timescale fluctuations in the prompt emission from many GRBs, and strong flares may emerge in their wake (e.g. Falcone et al. 2006). The shock breakout of a supernova in a probable Wolf-Rayet star has shown soft X-ray emission over about 2000s and UV emission over about 30 ks timescale (Campana et al. 2006). Cygnus X-3 has been going for years with high-energy photons, at a luminosity which would strongly affect any terrestrial-type planet within a few pc. This is of course improbable, given the number of objects such as Cygnus X-3 known, but it illustrates the usefulness of these results to estimate the effect of any new candidate objects of interest for terrestrial effects. Our results make possible a simplification in understanding the atmospheric chemistry effects of various kinds of ionizing photons. The ultimate atmospheric effect will depend upon the integrated fluence, and the short-timescale variability need not be included in estimating atmospheric effects on terrestrial planets.

Two caveats must be added to this independence claim. While the long-term atmospheric $O_3$ depletion may be nearly independent of burst duration, some of the biological impact may not be. The major part of the $O_3$ depletion timescale seen here from a nearly impulsive burst is of order 6 hours (while there may be a more gradual change over the previously mentioned weeks timescale). Thus, while the long-term atmospheric effects are nearly independent of duration,, the biological effects may worsen for events exceeding $10^4$ s, since a greater fraction of the UVB present in the event or generated in the upper atmosphere could reach the ground in a partially ozone-depleted atmosphere. Any post-burst flares (e.g. Flacone et al. 2006) which occur after this time and contain UVB would impact an already $O_3$-depleted atmosphere, and more UVB would reach the ground from the burst itself, at least on one side of the planet. For long events such as nearby supernovae, the ozone depletion develops more gradually, and may not reach full intensity for months. This can be more fully understood from appendix A. The independence from flux is true for the long-term depletion. However, for longer events the depletion at the end of the irradiation period is larger, even if the total fluence is the same. But the shorter events with the same fluence will, by the time the longer event has ended, typically have caught up. Aside from this, for a wide range of durations, given an assumed integrated spectrum, one need only know the integrated fluence in order to estimate the long-term atmospheric effects

4.2 *Effect of spectral variation*
For a peak energy from about 2 *keV* up to about 20 *MeV*, the atmospheric perturbation increases (as measured by ozone depletion), with effects nearly constant for even greater peak energy. This reflects the nearly constant attenuation rate of photons above a few *MeV* in the atmosphere. When our peak is at 20 *MeV*, a very small proportion of the flux is below 2 *MeV*. Higher energy photons are attenuated less, and penetrate more deeply before interacting. As seen in the figures, this causes more odd nitrogen production in the stratosphere, where ozone is most abundant. The effect is significant, approximately doubling the DNA-damaging UVB load over our range in peak photon energy.

One might anticipate stronger energy dependence. It does not exist for several reasons. Ozone is present over a range of altitudes, is not a constant fraction of the atmosphere (see Fig. 1), and due to the approximately exponential overall mass density profile of the atmosphere increased photon energy makes only a moderate change in the penetration depth. Also, $O_3$ depletion varies much less than linearly with increased irradiation, as noted in Thomas et al (2005b). There are three reasons for this: (a) The rate-limiting reaction (3) is more significant if more NO is present. (b) If more ozone is depleted, there is less available to deplete (c) Higher irradiation may lead to more ozone synthesis, which makes the net depletion less. We are plotting the overall ozone mass, and as seen in Fig. 4, for increased photon energy there is increased *synthesis* of ozone at lower altitudes (by solar UV), and this offsets the depletion effect somewhat. We include it in our overall results because it is still a part of the relevant optical depth for UVB to reach the surface.

Detailed modeling of a candidate event requires spectral information. One can make some generalizations: softer events such as X-ray flares from the Sun, dominated by low *keV* photons, are much less damaging for the same fluence than harder events. In particular, SHBs, which have not been included in studies of terrestrial planet effects, may acquire a new significance: they have a harder spectrum than long GRBs, and new data suggest their rate at low redshift is higher than formerly thought (Guetta & Piran 2006; If as consistent with present data their energies are two orders of magnitude lower and their rate two orders of magnitude larger than LSBs, they are of competitive expected fluence in an effectively thin disk galaxy. It will be important to use emerging new estimates of the SHB rate and improved spectral information to assess whether they may be competitive with LSBs as an irradiation hazard for terrestrial planetary biospheres. Data from GLAST on the high-energy spectrum will be useful here. Regarding LSBs, described by the Amati (Amati et al. 2002, Amati 2006) relation, the typical peak photon energy of LSBs varies as the square root of the isotropic equivalent energy; therefore more luminous ones will usually be more damaging than implied by simple luminosity scaling. Also, early X-Ray flares with substantial fluence may offer an additional hazard if ozone depletion is already underway. Ordinary supernovae may be more damaging than assumed previously, as part of their fluence arrives in an ozone-depleted atmosphere.

5. ACKNOWLEDMENTS
This work was supported by the NASA Astrobiology: Exobiology and Evolutionary Biology Program under grant number NNG04GM14G, computing at the National Center for Supercomputing Applications, the KU General Research Fund, and an Undergraduate Research Award from the University of Kansas Honors Program, and for M.M. by DOE grant DE-FG02-04ER54790. The University of Kansas Theatre Department assisted in making some of this science available to a wider audience. We thank Charles Jackman and the referee John Scalo for helpful comments.

APPENDIX A: A HEURISTIC DERIVATION OF THE INDEPENDENCE OF LONG-TERM OZONE DEPLETION FROM BURST DURATION

This derivation simplifies the $O_3$ depletion to the dominant processes that take place during and after the irradiation period. It is done to illustrate why it is that when the irradiation energy is held constant, but its length varied, that $O_3$ depletion at the end of the irradiation period increases with its duration, but that at later times the the ozone depletion level reached is independent of the duration of the earlier irradiation period. The atmosphere is assumed to be initially dominated by its usual consitituents $N_2$ and $O_2$. The reactions which dominate each process are included here, except that NO is taken as a proxy for both itself and $NO_2$ in depletion of $O_3$ , as they are closely coupled in the catalysis. In this notation, [x] refers to the density of x. Rate constants $k_n$ are as given in Thomas et al (2005b). J is the radiation flux onto the atmosphere, and $\sigma$ a typical cross section for ionization and dissociation of $N_2$. Atomic nitrogen formed from dissociation quickly reacts to form NO:

$d[N]/dt = \sigma J [N_2] - d[NO]/dt$            1

$d[NO]/dt = k_{16} [O_2] [N]$            2

therefore

$d[N]/dt = \sigma J [N_2] - k_{16} [O_2] [N]$            3

this is soluble by elementary methods,

$[N] = (\sigma J [N_2])/( k_{16} [O_2]) - C_1 \exp(- k_{16} [O_2] t)$            4

Where $C_1$ is a constant reflecting a transient we set equal to zero since N is normally close to nonexistent in the atmosphere, and we have a "steady-state" [N] which reflects the ratio of synthesis to loss, valid during the burst.

Substituting the resulting value of [N] into (2), we have

$d[NO]/dt = \sigma J [N_2]$            5

reflecting the fact that NO production is controlled by the flux, assuming that essentially all the N reacts this way, and that NO sinks (particularly rainout) have not yet become important.

Ozone depletion can be described by

$d[O_3]/dt = -[O_3][NO] k_7$            6

Substituting $\sigma J [N_2] t$ from (5) for [NO] into 6, we have

$$d[O_3]/dt = -[O_3] k_7 \sigma J [N_2] t \qquad 7$$

implying

$$[O_3] = [O_3]_0 \exp(-0.5 k_7 \sigma J [N_2] t^2) \qquad 8$$

But the above is only valid during the period of irradiation, which we shall call $t_1$. The total fluence is $F = J t_1$, and we can write the ozone abundance at the end of this period as

$$[O_3]_{t_1} = [O_3]_0 \exp(-0.5 k_7 \sigma [N_2] F t_1) \qquad 9$$

Now we calculate the continuing ozone depletion with the existing NO, taken as constant before rainout becomes important. From (5), the NO is frozen at the end of $t_1$ as

$$[NO] = \sigma J [N_2] t_1 = \sigma [N_2] F \qquad 10$$

Using this value in (6) for the post-irradiation period, we have

$$d[O_3]/dt = -[O_3] k_7 \sigma [N_2] F \qquad 11$$

which has the solution, using appropriate boundary conditions, evaluated at some later time $t_2 > t_1$,

$$[O_3]_{t_2} = [O_3]_{t_1} \exp(-k_7 \sigma [N_2] F (t_2 - t_1)) \qquad 12$$

Substituting (9) in (12),

$$[O_3]_{t_2} = [O_3]_0 \exp(-0.5 k_7 \sigma [N_2] F (t_2 - 0.5 t_1)) \qquad 13$$

From (9), the amount of depletion at the end of the irradiation period $t_1$, increases with $t_1$. This can be verified by a careful examination of Figure 7. For times $t_2 \gg t_1$ the depletion is essentially *independent* of $t_1$. This condition can be examined in our numerical results and verified for all but our $10^8$ s burst, which is the primary motivation of this derivation. For irradiation times longer than a few years, rainout begins to be important, which is a sink for NO, and the size of effects will obviously be much smaller.

These results can be understood much more simply, given the assumptions made. If basically all N that is produced goes into NO, and NO has no sinks except for rainout (which starts much later), then the amount of NO produced will reflect the total fluence. The mean NO concentration during the irradiation period will be the same for any such period, but for a longer period it will have more time to deplete $O_3$, which explains the increase in depletion during the burst for longer bursts. As NO is a catalyst, which depletes $O_3$ without being consumed, the ultimate long-term depletion reached will depend on the fluence alone, until rainout begins to be important at very long times.

FIGURE CAPTIONS

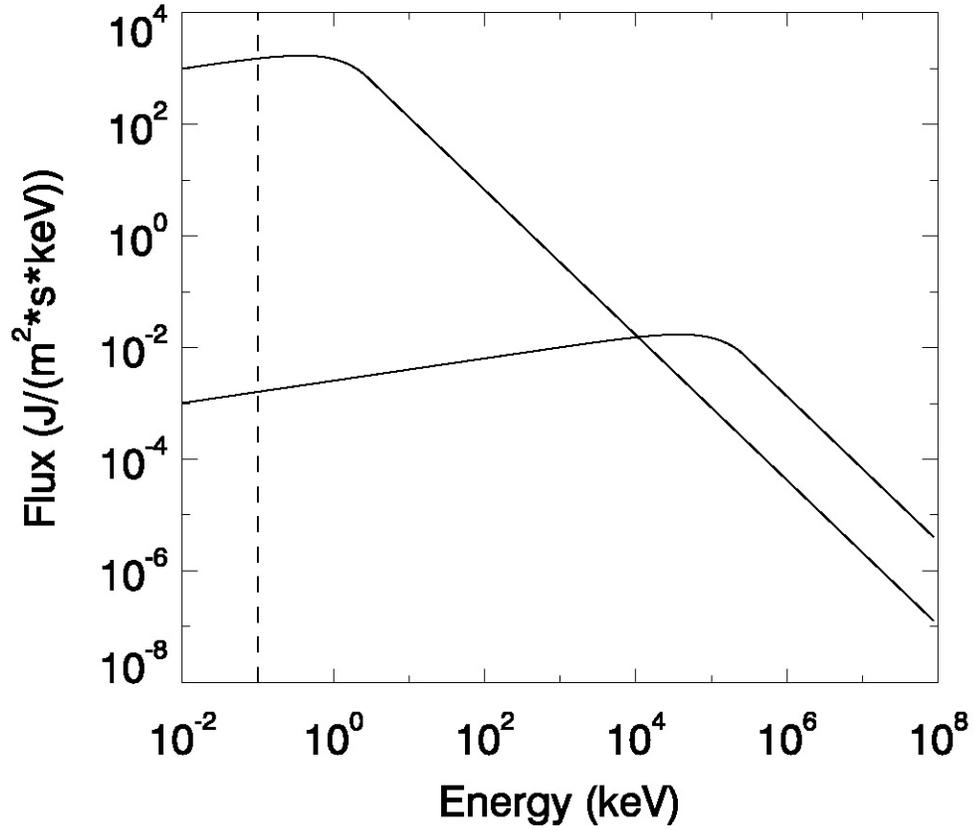

Figure 1: A plot derived from the Band spectrum for our lowest photon energy run, in which the peak energy parameter is set to 1.875keV, and the highest in which it is set to 187.5 MeV. Note that the units are energy flux, not photon flux, as is customary in the GRB literature, so an extra power of E is present. In the first case, about 1% of the energy is lost below our lower limit (vertical dashed line at 100 eV) and in the second, our worst-case about 10% of the energy is lost above our upper limit (RHS axis at 100 GeV).

# O₃ Density

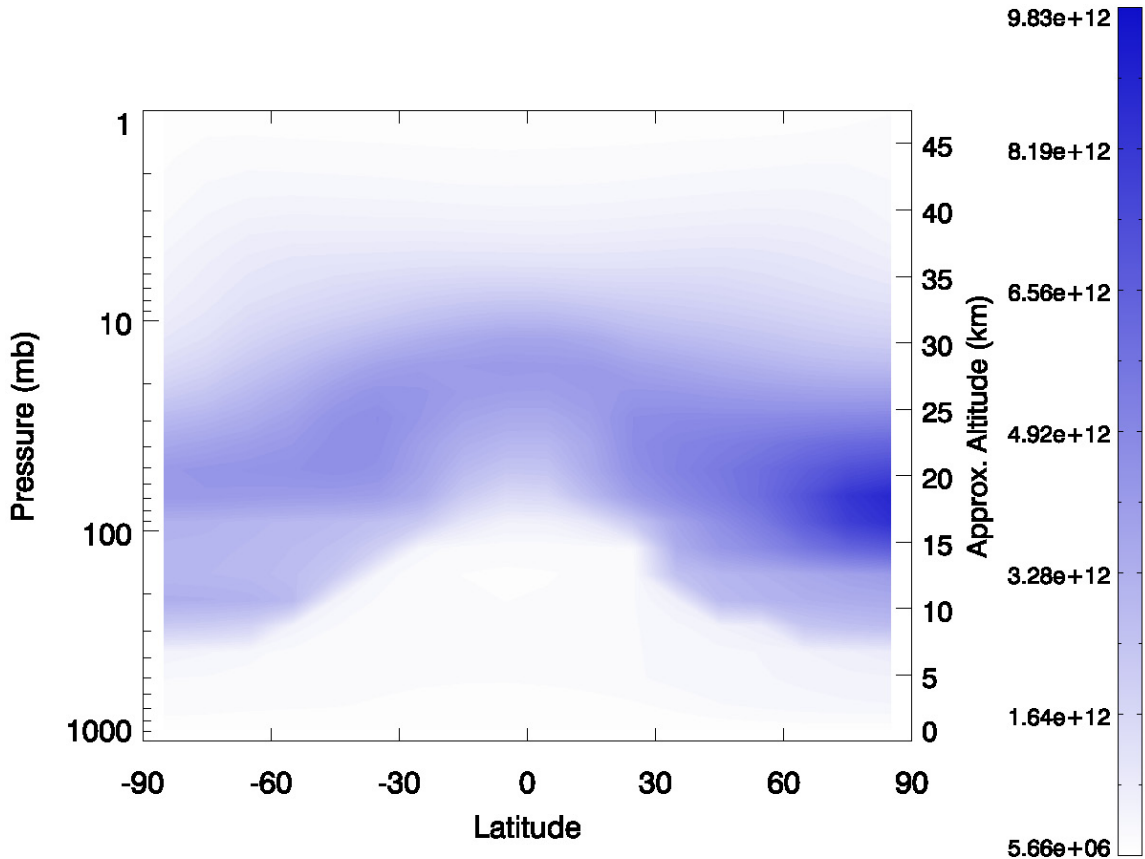

Figure 2: A plot of the latitude, log-pressure (approximately altitude) distribution of ozone density in cm$^{-3}$ in the unperturbed atmosphere. Most of the ozone is in the stratosphere, and lies between 10 and 35 km. Most GRB photons reach only the upper stratosphere, with only small numbers of interactions in the lower stratosphere. Variation of effect with energy shown later depends on the variation of penetration with photon energy.

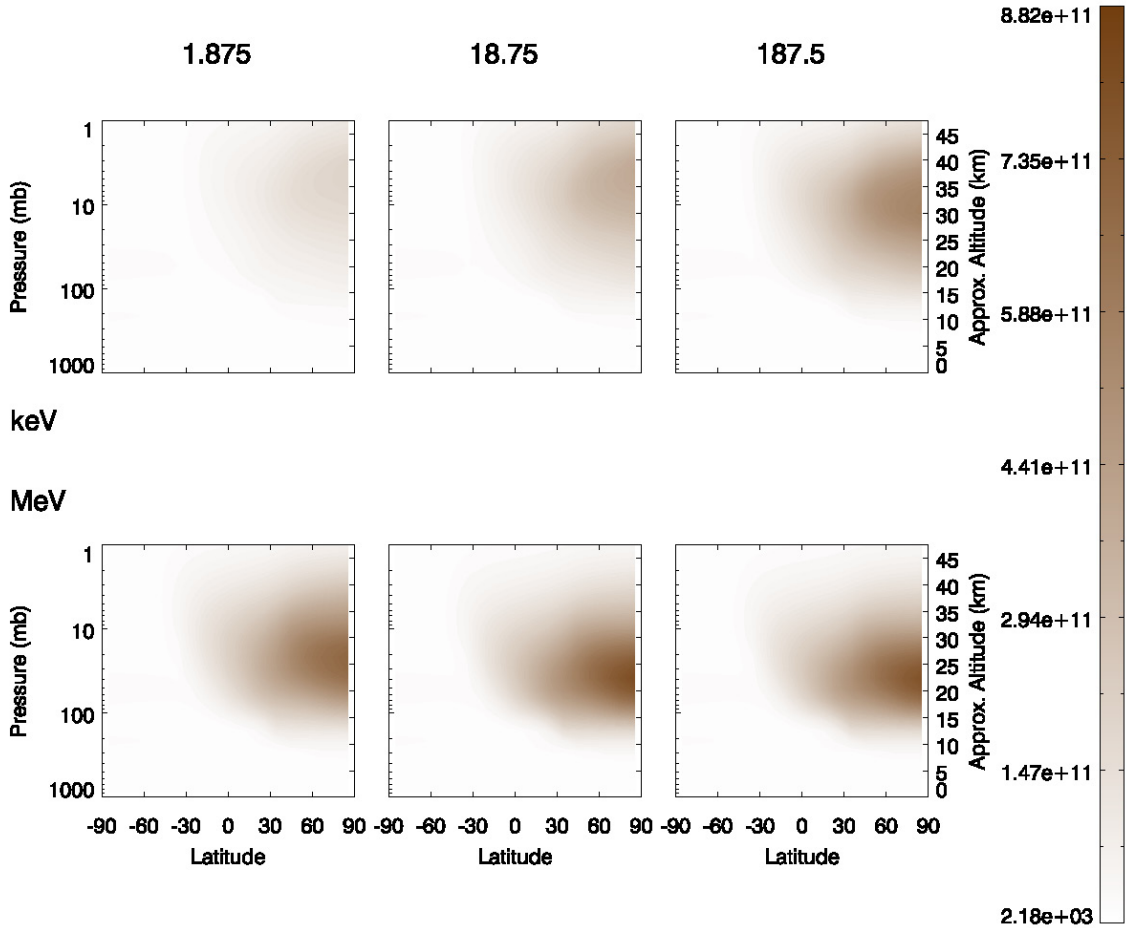

Figure 3: The latitude, log pressure distribution of the density in cm$^{-3}$ of odd nitrogen compounds five days after (10 s) bursts at 45° N with peak photon energies over six orders of magnitude, as described in the text. Higher energy photons penetrate deeper into the stratosphere, where the compounds produced are more protected from early destruction by photolysis.

# % δ O₃ Density

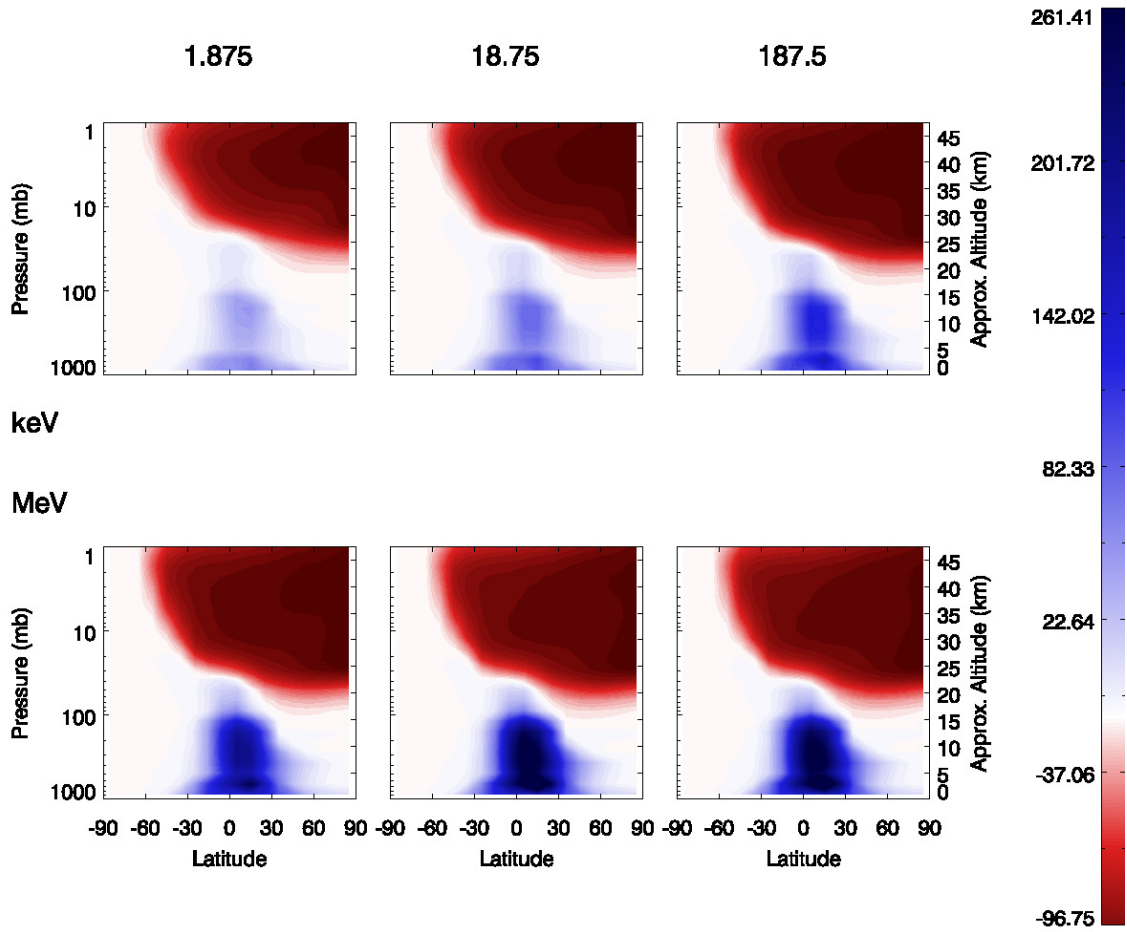

# % δ O₃ Density

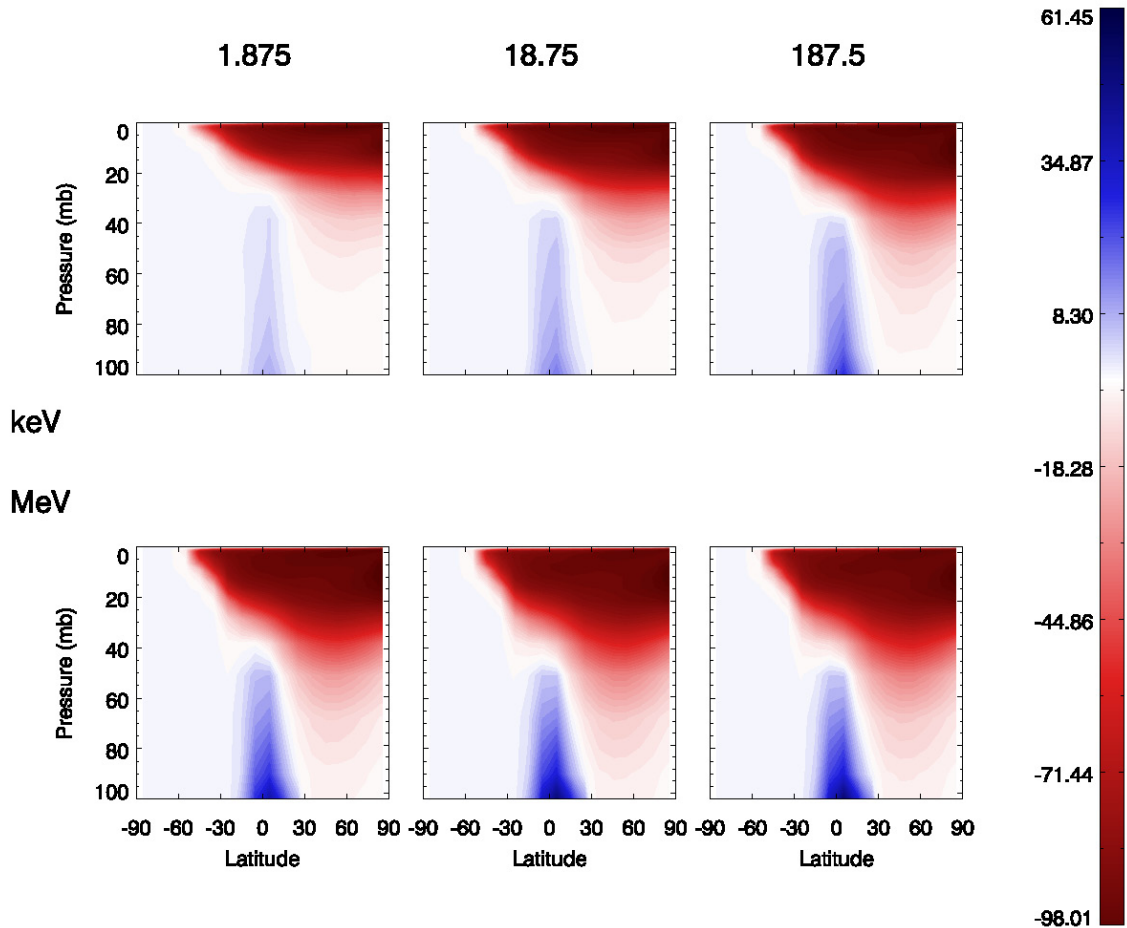

Figure 4: (a) The *percentage* change in ozone density produced by bursts of various energies, 11 days post-burst. Note that the isolated blob at the bottom of each panel represents a net ozone increase, while the upper part represents depletion. (b) The same information as plotted (a), except that the pressure axis is linear. This shows that although the difference in altitude of the boundary of depleted ozone is not great, there is a substantial variation in the mass of depleted ozone with increased photon energy.

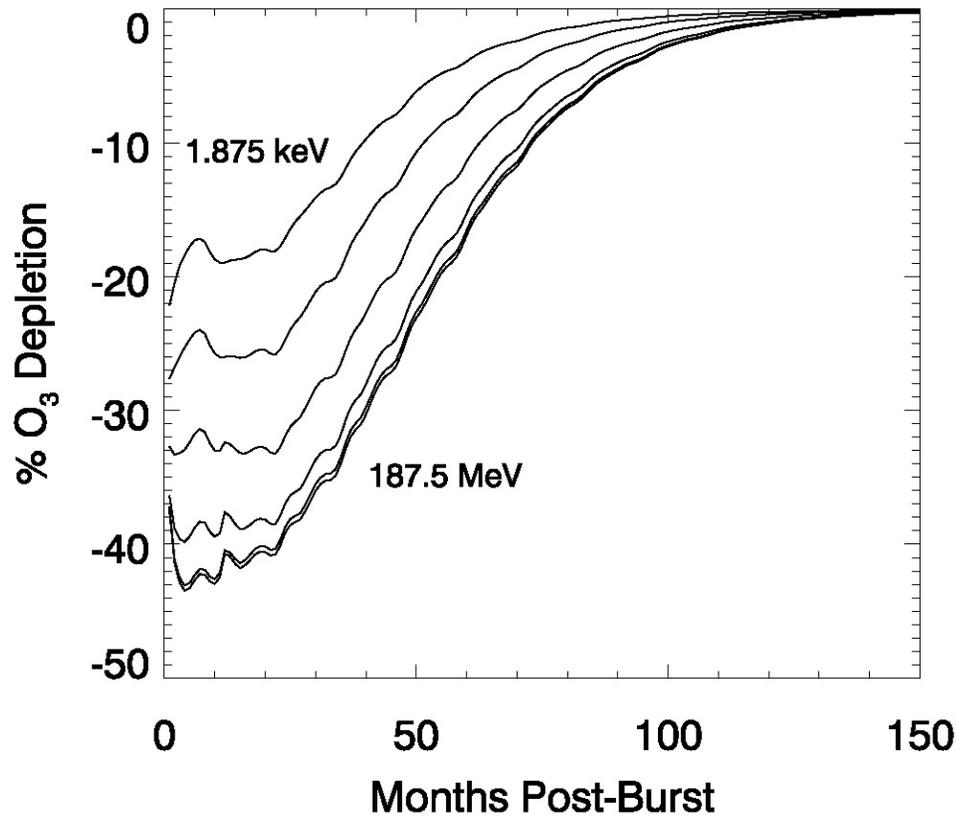

Figure 5: The global mass of ozone in the atmosphere as a percentage of its unperturbed value, as a function of time after the burst, for various peak photon energies. The lines are not labeled, but the depletion percentage increases monotonically from our lowest (1.875 keV) to highest (187.5 Mev) peak photon energies, which vary by one order of magnitude per line. The bottom two lines nearly overlie one another. The effect of depletion with increased photon energy is partly compensated by ozone synthesis in the lower stratosphere.

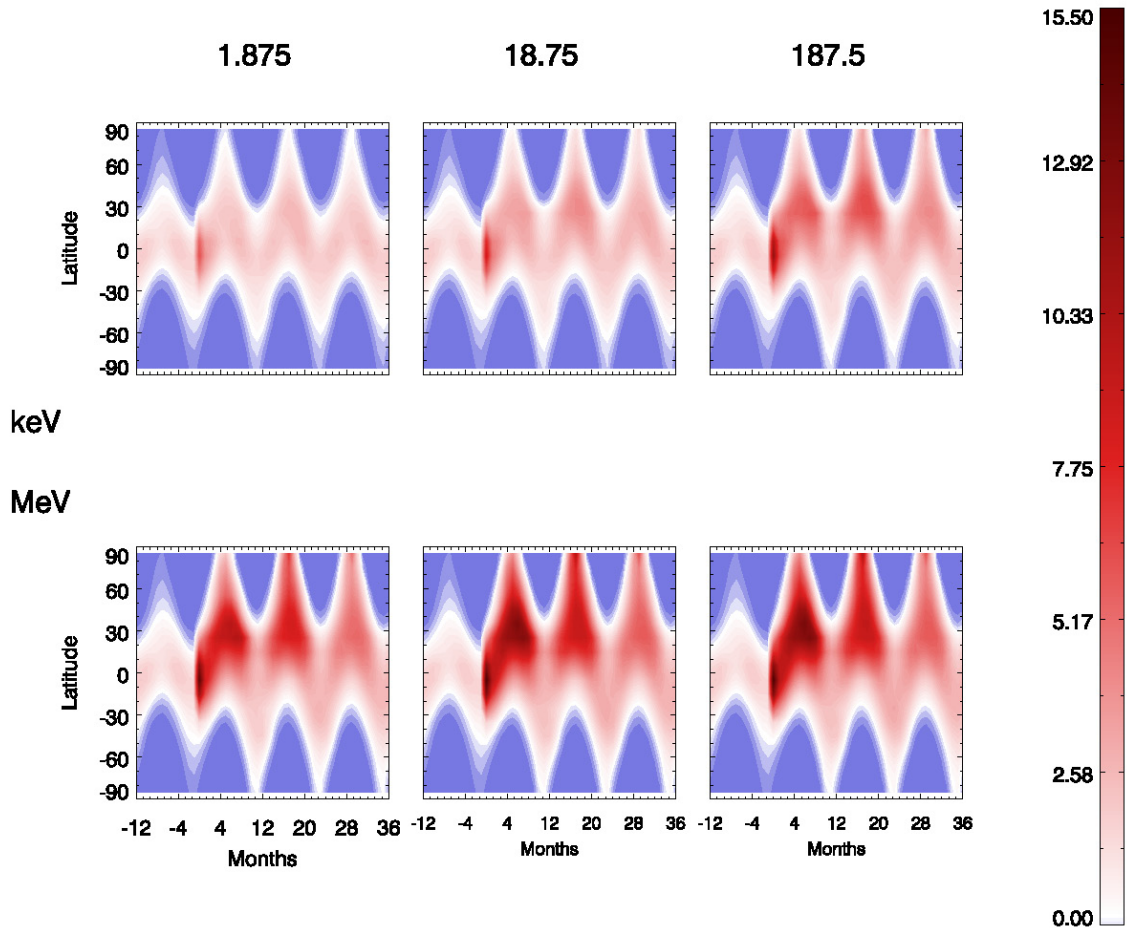

Figure 6: The solar UVB which is transmitted given reduced optical depth for given ozone depletion is convolved with a biological weighting function for DNA damage, the change in optical depth and total irradiation as a result of the incidence angle of the sun and the length of the day, to get an estimate of total UVB damage as a function of latitude and time for each of our burst peak photon energy cases. The annual oscillations are merely due to the apparent north-south movement of the Sun in the sky, which changes the latitude of most nearly normal Sun incidence, the length of the day, etc. The plots are normalized to white, the pre-burst annual global mean of this DNA damage measure, and one year pre-burst is included for comparison. Severe damage is indicated in most cases when this ratio exceeds approximately 2.5 (see Thomas et al. 2005b for discussion). The damage increases for peak photon energies up to about 20 MeV, due to more efficient ozone destruction, then it levels off

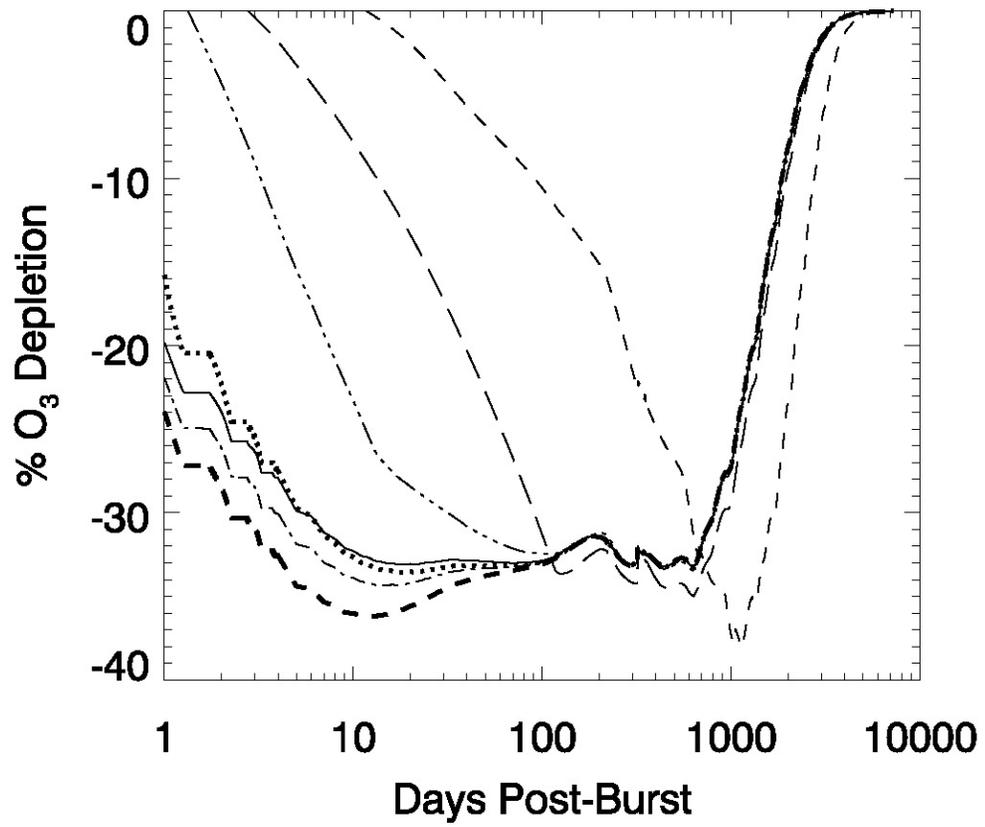

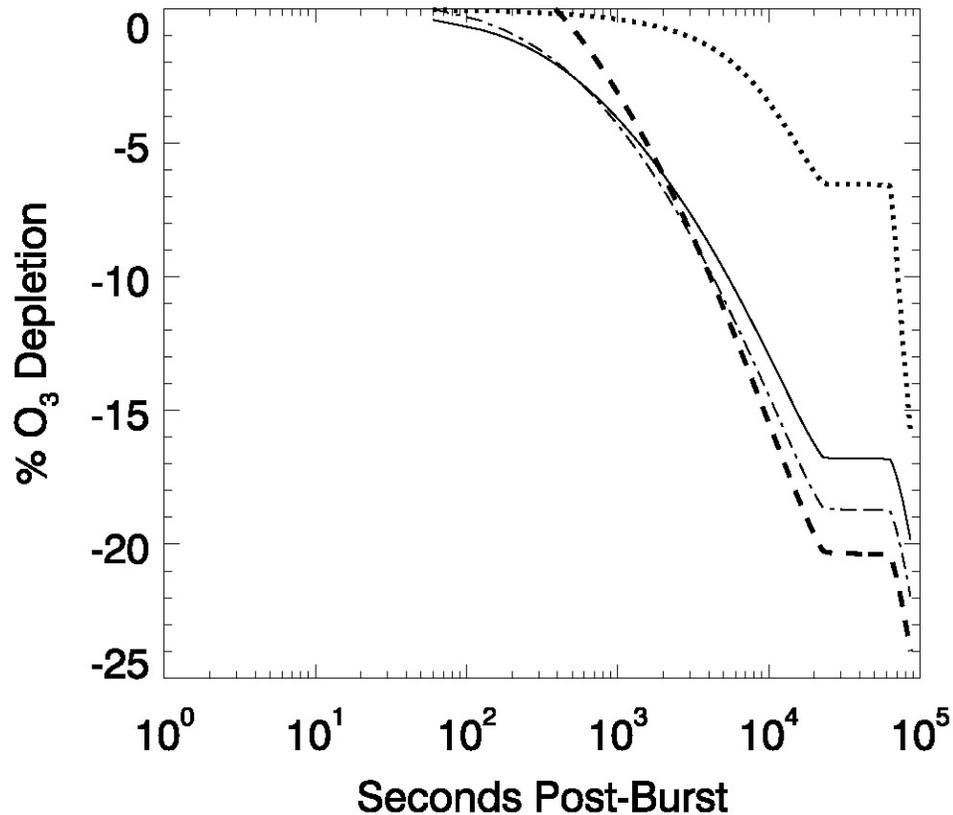

Figure 7: (a) The fractional global ozone mass (relative to its unperturbed value), as in Figure 5, as a function of time for the 187.5 keV case. The energy is deposited uniformly over times $10^{-1}$ s (thick dash), 10s (dot-dash), $10^3$ s (solid), $10^5$ s (dotted), $10^6$ s (dash-3dot), $10^7$ s (long dash), and $10^8$ s (short dash). Maximal global ozone depletion varies only slightly, though onset is obviously delayed for long events. The longest events cause the greatest depletion as they are able to delay recovery by producing odd nitrogen compounds which would have been destroyed if produced early by a short burst, as well as include transport effects. Slope discontinuities visible at early times here and in part b correspond to changes in photolysis reactions at sunset and sunrise in the stratosphere during the day (the burst is assumed to have taken place at noon). Oscillations on 100s of days timescale are seasonal. (b) As in part a, except that values are plotted only for the first day for those bursts which have durations of one day or less.